# Water vapour pressure as determining control parameter to fabricate high efficiency perovskite solar cells at ambient conditions

Lidia Contreras-Bernal[b], Juan Jesús Gallardo,[b] Javier Navas,[b] Jesús Idígoras[a],*, Juan A. Anta[a],*

[a]*Área de Química Física, Universidad Pablo de Olavide, E-41013, Sevilla, Spain*
[b]*Departamento de Química Física, Facultad de Ciencias, Universidad de Cádiz*, E-11510 Puerto Real
(Cádiz), Spain

**Abstract:** Although perovskite solar cells have demonstrated impressive efficiencies in research labs (above 23%), there is a need of experimental procedures that allow their fabrication at ambient conditions, which would decrease substantially manufacturing costs. However, under ambient conditions, a delicate control of the moisture level in the atmosphere has to be enforced to achieve efficient and highly stable devices. In this work, we show that it is the absolute content of water measured in the form of partial water vapour pressure (WVP) the only determining control parameter that needs to be considered during preparation. Following this perspective, MAPbI$_3$ perovskite films were deposited under different WVP by changing the relative humidity (RH) and the lab temperature. We found that efficient and reproducible devices can be obtained at given values of WVP. Furthermore, it is demonstrated that small temperature changes, at the same value of the RH, result in huge changes in performance, due to the non-linear dependence of the WVP on temperature. We have extended the procedure to accomplish high-efficient FA$_{0.83}$MA$_{0.17}$PbI$_3$ devices at ambient conditions by adjusting DMSO proportion in precursor solution as a function of WVP only. As an example of the relevance of this paramater, a WVP value of around of 1.6 kPa appears to be an upper limit for safe fabrication of high efficiency devices at ambient conditions, regardless the RH and lab temperature.

**Introduction**

The photovoltaic field has undergone rapid progress in the last few years due to the development of solar cells based on hybrid organic-inorganic perovskite materials.[1,2] The power conversion efficiency (PCE) of perovskite solar cells (PSCs) is evolving far faster than any other solar technology known to date.[3–5] The natural abundance of the raw materials employed to synthesise perovskite and their excellent optoelectronic properties (large absorption coefficient, direct band gap, high charge carrier mobility and long carrier diffusion lengths) make them potential competitors of well-established thin-film photovoltaic technologies such as those based on silicon.[6–8] Indeed, nowadays, the PSCs can be fabricated with competitive efficiency (PCE > 20 %) [5,9–13], setting the certified record in 23.7%.[14] However, these high PCEs have been achieved using testing-devices with a small active area and fabricated under dry conditions inside a glove-box (GB). These two issues do not only increase the manufacturing cost and limit its industrial scale, but also restrict its commercialisation.[15–17]

Although perovskite films under certain moisture level have been achieved with a high quality,[18–21] the fact is that these devices have not reached yet the PCEs of their glove-box counterparts.[22] Nevertheless, different strategies have been recently reported in the literature to optimize PSCs fabrication under ambient condition.[23–27] One of the most recent methodology is that proposed by Aranda *et. al.*[26] They demonstrated that it is possible to prepare highly efficient methylammonium lead iodide (MAPbI$_3$) perovskite devices under different humidity conditions by controlling the dimethyl sulfoxide (DMSO): $Pb^{2+}$ ratio in the precursor solution as a function of the relative humidity (RH) of the environment. Using this methodology, PCEs approaching 19% for devices fabricated in humid environment were achieved, which are in the range of the highest reported efficiencies for this type of perovskite prepared in dry conditions.[1,28] In addition, these PSCs fabricated at ambient conditions have proven to be more robust against moisture-induced ageing and degradation than those prepared at low RH values or dry conditions.[29] The addition of both thiocyanate ions or orthosilicate in perovskite precursor solution[24,25,30] or the use of pre-heating treatments of the electron selective contact[23,27] have been suggested as other strategies to fabricate PSCs with high efficiency at ambient conditions.

In all previous reports related to the preparation of PSCs outside GB, RH is considered as the main control parameter to fabricate solar devices. Nevertheless, RH is a relative parameter that depends of the saturation value of water vapor in air, which depends on the temperature. The maximum amount of water molecules in the vapor phase (saturation), which is able to hold a system is well established in engineering and thermodynamics tables. These data are approximately determined by the Clausius-

Clapeyron equation

$$\ln P_{ws} = -\frac{\Delta H_V}{RT} + C \qquad (1)$$

where $P_{ws}$ and $\Delta H_v$ are the saturation vapour pressure and the water vaporization enthalpy, respectively. $T$ is the temperature, $R$ is de ideal gas constant and $C$ is a constant. Considering *Eq.*(1), a larger vapour pressure is expected for higher temperatures with an exponential correlation. Therefore a small temperature change brings about a larger modification of water content in air, measured as their partial vapor pressure $P_w$.[31] RH is actually defined as[32]

$$R.H\ (\%) = \frac{P_w}{P_{ws}} \cdot 100 \qquad (2)$$

Therefore, RH is not an indicative of absolute water content in air which is intimately related to the temperature of the system via *Eq.* (1). This could be the reason of the broad spread of PCE found in the literature for same type of perovskite prepared at the same values of RH, especially taking into account that a small temperature variation produces a huge change in $P_w$.[25,26] Therefore, RH is not by itself sufficient as control parameter to study and fabricate PSCs under atmospheric conditions. The ambient temperature needs to be considered as well. In this context, it has been recently proposed to use the dew point as a control parameter in the fabrication procedure.[33] Playing with several temperatures at the same RH, these authors have found an optimum value of the dew point in which the PCE was highest, showing that it is the absolute content of water the only parameter that matters. However, they only took into account the moisture levels during the annealing stage ignoring the fact that the water content in air affects the perovskite crystallization process in other stages of perovskite films fabrication as well.[20] In this work we choose to use the water vapour pressure (WVP) only as control parameter because it has a more direct and clear physical meaning associated to the RH, temperature and vaporization enthalpy via Eqs. (1) and (2).

Herein, we analysed the impact of WVP during the fabrication process of perovskite devices at ambient conditions in order to achieve high-efficiency solar devices. For that purpose, FTO/c-TiO$_2$/m-TiO$_2$/perovskite/Spiro-OMeTAD based devices have been fabricated using MAPbI$_3$ perovskite precursor solution with different DMSO:Pb$^{2+}$ ratio.[26] This perovskite precursor solution was deposited by spin-coating at different environmental conditions (outside GB). In particular, MAPbI$_3$ based devices were fabricated at different temperatures and RH values. UV-Vis spectra, photoluminescence (PL) measurements and impedance spectroscopy (IS) analysis were employed as characterization techniques to explain the photovoltaic parameters obtained for the PSCs as function of WVP. Finally, by setting WVP as the main control humidity parameter, we have achieved $FA_{0.83}MA_{0.17}PbI_3$ based PSCs under highly humid conditions. To our best knowledge, this is the first work in the literature where the photovoltaic parameters are reported as a function of the absolute content of water molecules in air in the form of a water vapour pressure.

**Results and discussion**

Figure 1 shows the best current-voltage curves obtained for $MAPbI_3$ devices prepared at the same value of the RH (50 %) but different temperatures (298 K, 299 K and 301 K). In accordance to *Eq.* (1), higher temperature implies a larger value of the water content in the laboratory atmosphere. In particular, WVP values of 1.58, 1.72 and 1.85 kPa were obtained from engineering tables (approximated by Eq. (1)) and Eq. (2) for those environmental conditions, respectively. PCEs of 16.7 ± 0.6 %, 11.4 ± 0.6 % and 11.3 ± 0.9 % were obtained at 1.58, 1.72 and 1.85 kPa, respectively. The statistical data of characterization photovoltaic parameters are shown in Figure S1 (Supplementary

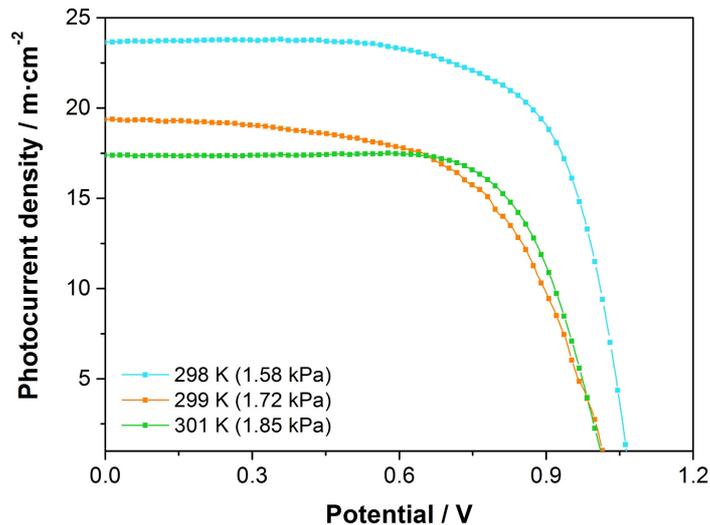

Information) as a function of WVP.

**Figure 1.** Current–voltage curves of $MAPbI_3$ perovskite devices prepared under different temperature conditions but same relative humidity (50 %) (resulting in different WVPs, as indicated). The curves have been measured in reverse scan under 1 sun—AM 1.5 illumination and using a mask of 0.16 $cm^2$. For all three cases, perovskite films were deposited from a solution precursor with a ratio DMSO:$Pb^{2+}$ of 0.75

It should be noted that similar PCE values were previously reported for MAPbI$_3$ devices deposited at 50% RH although no specific temperature data was given.[25,26] In this work we show, though, that small temperature changes have a strong impact on the performance, as a consequence of the big jump in the absolute content of water in air.

Although the perovskite films looked homogenous regardless of the environmental conditions during the deposition process, a drop of photocurrent density at short-circuit ($J_{sc}$) and open-circuit potential ($V_{oc}$) was clearly observed with the increase of temperature and, consequently, with higher WVPs (Fig. S1.B and Fig. S1.C). Therefore, these results plainly demonstrate the importance of WVP, as the main control parameter to take into account humidity, instead of RH as popularly used in PSC fabrication.

To analyse the effect of DMSO in perovskite precursor solution as a function of water content in air, precursor solutions with different DMSO:Pb$^{2+}$ ratios were employed. WVPs of 1.66 kPa and 1.85 kPa (301 K and RH of 44 % and 49 %, respectively) were considered in these experiments. Figure S2 shows the average photovoltaic parameters obtained using different DMSO:Pb$^{2+}$ ratios in the precursor solutions. The optimum DMSO:Pb$^{2+}$ ratio for the different WVP are in line with the mechanism previously discussed.[26,29] Under ambient conditions, the atmospheric H$_2$O molecules can be incorporated to the lead complex and compete with the DMSO in the coordination sphere of the lead ion. Thus, in humid atmospheres, it is necessary to adjust the content of DMSO in the perovskite precursor solution as a function of WVP to achieve the correct stoichiometry of the PbI$_2$:H$_2$O:DMSO complex. In general, lower amounts of DMSO are required to reach high efficiencies at higher WVP. In particular, an optimum DMSO:Pb$^{2+}$ ratio of 1.0 and 0.75 were found for 1.66 kPa and 1.85 kPa, respectively, for which a maximum efficiency of 15.6 % and 12.5 % were obtained (Fig. S2.A). Nevertheless, in both cases, the values obtained using the different DMSO:Pb$^{2+}$ ratios at these environmental conditions were mainly determined by the distinct $J_{sc}$ values (Fig. S2.B). In contrast, no significant differences were found for $V_{oc}$ values (Fig. S2.C). A further analysis is presented below.

On the other hand, regardless the DMSO:Pb$^{2+}$ ratio, lower efficiencies were revealed for the highest WVP (Fig. S2.A). These results are line with the efficiency trends previously reported in which only the RH values were taken into account as main parameter (Fig. 1).[19,26] For atmospheric conditions of 1.85 kPa, lower efficiencies (around 12 % for ratio 0.75) were obtained as compared with those reported for the same perovskite deposition methodology (around 15 % for ratio 0.75).[26] This worse performing device could be due to the PSCs were here fabricated at a higher temperature (301 K) than that usually reported in literature.[19,27,33] Thus, although the RH were the same, the content of water in air (WVP) has been higher in our experiments than any other before. In this context, a drop of photovoltaic parameters was observed for 1.85 kPa with respect to 1.66 kPa, except to the FF that showed a similar value for intermediate ratios (Fig. S2.B, Fig. S2.C and Fig. S2.D). This further demonstrates that it is the absolute water content in air what really matters when the perovskite film is fabricated.

**Figure 2.** Efficiency obtained for MAPbI$_3$ devices as a function of the water vapor pressure (WVP) during perovskite deposition. Alternative values of the DMSO:Pb$^{2+}$ ratio of the precursor are compared

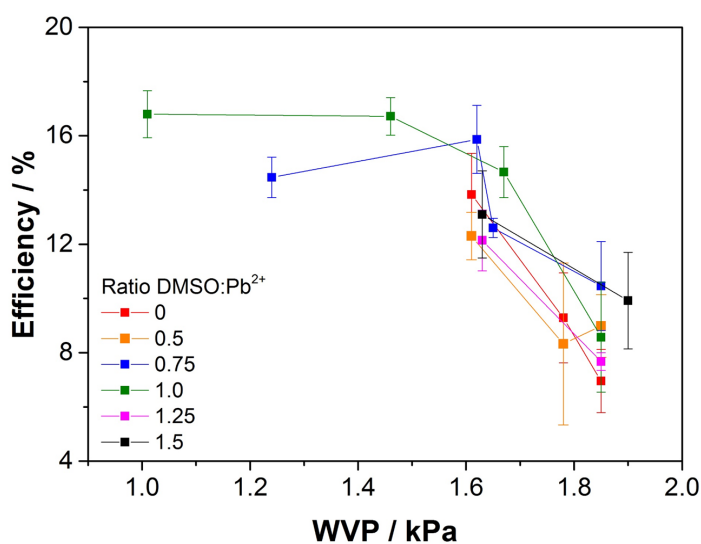

as indicated. The efficiency has been extracted from current-voltage curves measured under 1 sun—AM 1.5 illumination using a mask of 0.16 cm$^2$.

Average efficiencies for a large range of WVPs are shown in Figure 2. The range of WVP analysed corresponds to typical atmospheric conditions throughout one year in a region with Mediterranean climate. The highest efficiencies were found for WVP values lower than 1.66 kPa. Specifically, 16.8 ± 0.9 % efficiencies were obtained for devices prepared at 1.06 kPa (301 K and 28 % RH). In this regime (from 1.0 kPa to around 1.6 kPa), the performance of devices seems to be insensitive to the content of water in air as shown by the plateau obtained for the optimum DMSO:Pb$^{2+}$ ratio of 1.0. The maximum efficiency found in this work (18.2 % for 1.06 kPa) is in line with the maximum reported PCE for MAPbI$_3$ devices.[1,26,28] However, WVPs beyond 1.7 kPa lead to a strong drop in efficiency. This is the case of 1.85 kPa (corresponding to 301 K and 49

% RH) in which the performance of the devices drops to 10.5 ± 1.6 % using even the optimum DMSO:$Pb^{2+}$ ratio of 0.75. Previously, it has been reported that low performance of PSCs at high temperature could be due to a low solubility of perovskite materials in the precursor solution, which could affect the growth of perovskite crystals.[34,35] Nevertheless, it should be noted that the same high temperature was here reached to both the lowest and highest WVP analysed. Therefore, the dispersion in efficiencies appears to result from effects of atmospheric water content during the PSCs fabrication but not due to a change of perovskite solubility. This is an extra argument to further support the importance of using the absolute content of water as control during the perovskite deposition process

On the other hand, we also observed that the stability under illumination and the reproducibility were worse for the bad performing devices. In consequence, a WVP value of around of 1.6 kPa appears to be an upper limit for safe fabrication of high efficiency devices at ambient conditions.

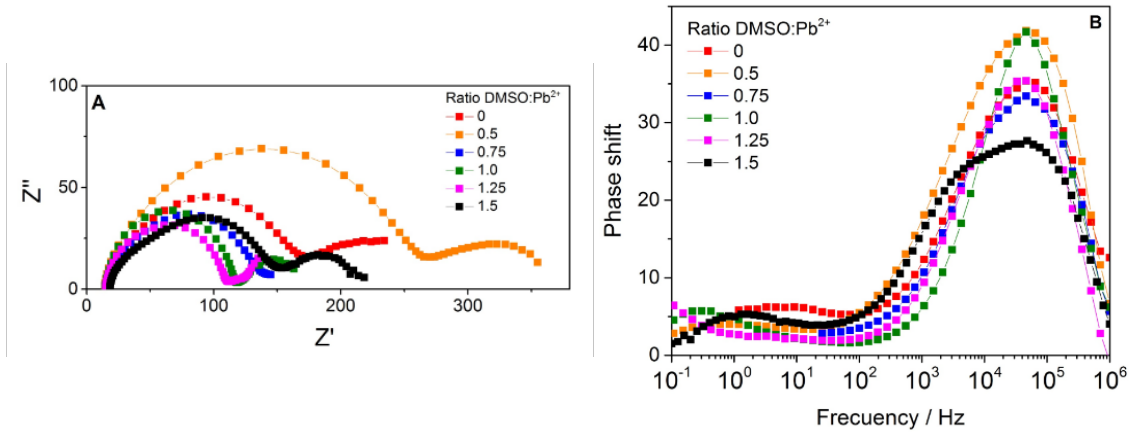

**Figure 3.** Impedance Nyquist (A) and Bode (B) plots for MAPbI$_3$ devices prepared with different DMSO:$Pb^{2+}$ ratios at 1.66 kPa. Data obtained under red illumination (635 nm) are shown. The open-circuit photopotential generated by the illumination is 0.911 V.

As it has been shown above, the efficiency trends for a certain WVP are mainly governed by the photocurrent, whereas the $V_{oc}$ happens to be basically insensitive to the DMSO content in the precursor solution (Fig.S2). To analyse the impact of DMSO:$Pb^{2+}$ ratios on performing devices, impedance spectroscopy measurements at open circuit potential in the range of frequency of $10^6$-0.1 Hz were performed. Due to the lack of stability during impedance measurement shown by devices prepared under higher environmental WVP values, the impedance analysis was restricted to devices fabricated at 1.66 kPa. The measurements were carried out at two excitation wavelengths (465 nm and 635 nm) to establish the impact of different optical penetration lengths in the perovskite layer. Similar methodology was previously reported by us.[36–38] Figure 3.A and 3.B show the impedance response in the form of Nyquist and Bode plots, respectively, for various DMSO:$Pb^{2+}$ ratios recorded at 635 nm. The impedance spectra were mainly characterized by two kinetics signals, embodied as semicircles (arcs) in the Nyquist plot and as peaks in the Bode plot. The high frequency (*HF*) signal at $10^5$ Hz has been attributed to electronic transport and recombination processes in the perovskite layer[36,39–42] while the signal that appear at low frequency (*LF*) has been associated with ionic motion and charge accumulation at the contacts.[43,44]

The *HF* resistance can be extracted from the impedance data by fitting to an equivalent circuit. In this work, a simple Voigt circuit [$R_s$−($R_{HF}$·$CPE_{HF}$)] has been used to extract the corresponding *HF* resistance $R_{HF}$.[45] The $R_{HF}$ obtained for the two excitation wavelengths are plotted as a function of open-circuit potential originated by different illumination intensity in Figure S3. In particular, the $R_{HF}$ is found to vary exponentially with the open-circuit potential as predicted by the following equation[36,46]

$$R_{HF} = \left(\frac{\partial J_{rec}}{\partial V}\right)^{-1} = R_{00} exp\left(-\frac{\beta q V}{k_B T}\right)$$

where $J_{rec}$ is the recombination current, $R_{00}$ is the resistance at zero potential, $k_B$ is the Boltzman constant, $T$ is the absolute temperature, and $\beta$ is the transfer or recombination parameters. On the other hand, the ideality factor (*m*) is related to $\beta$ as $m=1/\beta$. Results for the *HF* resistance are in line with previous impedance analysis:[36,37] The ideality factors were close to 2 and unchanged with respect to the generation profile, which is associated with Shockley−Read−Hall recombination in the bulk of the active layer.[36,47,48] Activation energy for recombination of the employed perovskite materials was also determined (1.60 eV) by extrapolating the $V_{oc}$ for different temperature data to $T \rightarrow 0$[49] (Fig. S4.B). This activation energy was coincident with the band gap reported for $MAPbI_3$ perovskite[50] and with the optical band gap (Fig. S4.A) extracted from absorption data of the devices (1.63 eV), which confirms that the recombination is mainly determined by the bulk of the perovskite layer itself at open-circuit conditions. Besides, we found that the $R_{HF}$ was basically independent from the DMSO:$Pb^{2+}$ ratio, which explains the relative constancy of the $V_{oc}$ values with respect to the precursor composition (Fig. S2).

Nevertheless, when the $R_{HF}$ extracted for two different WVPs (1.66 kPa and 1.85 kPa) is compared, lower values of $R_{HF}$ for highest WVP but same ideality factor (close to 2) are obtained. This is shown in Figure S3.B. This indicates a clear net enhancement in the Shockley−Read−Hall recombination rate for 1.85 kPa, which explains the reduction of the open-circuit potential at 1 *sun* illumination reported in Figure 1, Figure S1.C and Figure S2.C. This means that too much water in the crystalline structure of the perovskite (above the upper limit of 1.6 kPa mentioned above) triggers more recombination in the bulk of the perovskite, probably due to the formation of more recombination centers.

To investigate the influence of DMSO:$Pb^{2+}$ ratio as a function of water content in air on both absorption and charge extraction/separation performance, steady-state photoluminescence (PL) and UV-Vis spectroscopy studies were conducted. These were done in perovskite films deposited on mesoporous $TiO_2$ at 1.66 kPa and 1.85 kPa, respectively and the results are presented in Fig. 4. PL data have been normalized to the maximum absorbance found for each device by dividing the whole UV-Vis spectrum by the maximum intensity and multiplying the PL peak by the factor thus obtained. This way, a better comparison of different perovskite films is ensured.

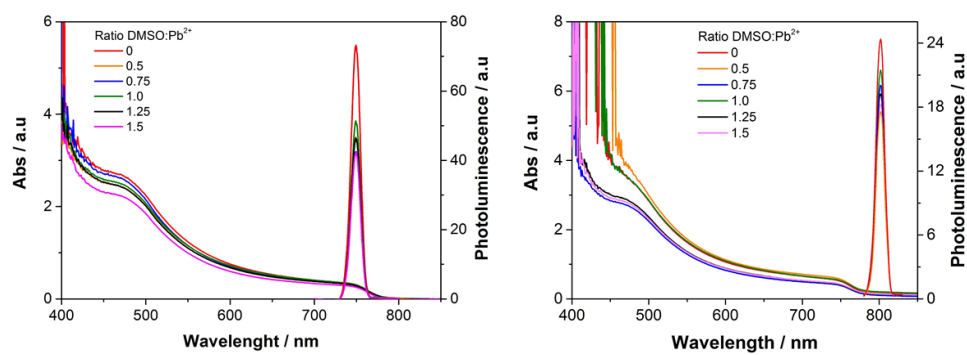

**Figure 4.** Optical absorption and normalized steady state photoluminescence spectra of MAPbI$_3$ films deposited on TiO$_2$ substrates from precursor solution with different DMSO:$Pb^{2+}$ ratios. Data obtained for two WVP: 1.85 kPa (left) and 1.66 kPa (right). Excitation wavelength of 497 nm and 532 nm were used, respectively.

From Figure 4, we infer that lower DMSO:$Pb^{2+}$ ratios produce perovskite films with higher absorption spectra at short wavelengths regardless the WVP used. This effect is more acute at 1.66 KPa. Bearing in mind that the perovskite films thickness are similar for all devices studied (~ 400 nm), the changes in absorbance could arise from changes in lead ion coordination, which depends on the mixture solvent in the precursor solution.[51] This is in line with the improved crystallinity observed for DMSO:$Pb^{2+}$ ≥ 1.0 by the increase in the X-ray diffraction (XRD) peak intensity at 14.1°, as shown in Figure S5. (This peak is associated with (110) plane of tetragonal $MAPbI_3$.) The XRD was carried out for perovskite films deposited at 1.01 kPa, for which a peak at 12.6° was also found for the different DMSO:$Pb^{2+}$ ratios. This peak is attributed to the (001) lattice plane of hexagonal $PbI_2$. A qualitative analysis of these peaks revealed that the highest $PbI_2$/$MAPbI_3$ ratio is achieved for films prepared at DMSO:$Pb^{2+}$ ratio ≤ 1.0 (Table S1). Bearing in mind that the highest efficiency was obtained for ratio 1.0 for low WVP, this result points out that a certain $PbI_2$ content on perovskite films could be the reason of the better performing devices.[47]

It is also interesting to analyse the impact of the ratio on both light harvesting and electron injection. In Figure 4 it is observed that the lowest DMSO:$Pb^{2+}$ led to the highest photoluminescence peak intensities. Since the PL signal intensity arise from the radiative recombination processes inside perovskite material, an increased PL signal is indicative of slower electron injection from the perovskite film to the $TiO_2$ layer. Therefore, we conclude that injection and absorption are somehow compensated, which explains the relative insensitivity of the photocurrent with respect to the DMSO:$Pb^{2+}$ ratio (Fig. S2.B). By contrast, there is a clear difference between the PL intensities revealed at different WVP. Faster charge extraction was obtained for the films prepared at 1.66 kPa than those prepared at 1.85 kPa. Furthermore, for 1.66 kPa, an increase of the absorbance at short wavelengths for lowest DMSO:$Pb^{2+}$ ratios is also observed. These results are in line with the drop in photocurrent obtained at higher WVP (Fig. 1, Fig. S1. B and Fig. S2.B).

Scanning electron microscopy (SEM) images show also a distinct perovskite morphology depending on the WVP and DMSO:$Pb^{2+}$ ratio (Fig. S6). The morphological change as function of DMSO content in precursor solution is in agreement with the data reported by Aranda et al.[26] For 1.66 kPa, perovskite films deposited from precursor solutions with low content in DMSO (ratio 0.5) revealed a fiber-like morphology mixed with larger crystalline domains (Fig. S6.E), while for ratio 1.0 these fibers did not show up, and a totally covered surface is observed (Fig. S6.C). Although, large domains were also formed for highest content in DMSO (ratio 1.5), non-homogeneous films were detected (Fig. S6.A) due to presence of pin-holes. Therefore, an optimized stoichiometry of $PbI_2$:$H_2O$:DMSO leads to a more adequate morphology, in line with the best performing devices. The same trend was observed for films fabricated at 1.85 kPa. However, a higher pin-hole proportion was revealed for the highest DMSO:$Pb^{2+}$ ratios (Fig. S6.B). Indeed, non-homogeneous films were already detected for ratio 1.0 (Fig. S6.D). On the other hand, smaller crystalline domains were observed when the films were prepared at higher WVPs. In particular, perovskite domains in the range of 200 nm and 100 nm were obtained for 1.66 kPa and 1.85 kPa, respectively. Considering that the grain boundaries increase the non-radiative recombination rate as well as adversely affect to charge extraction efficiency at the interface between perovskite and selective contact,[28,52] these results could explain the low $V_{oc}$ (due to the larger concentration of recombination centers, as stated above) (Fig. 1, Fig. S1 and Fig. S2) and the poor injection efficiency (Fig. 4) obtained for devices prepared under higher WVP.

In order to show the wide applicability of a preparation method based on the control of the absolute water content in air, we have attempted the preparation of mixed perovskites at ambient conditions. Thus, we have extended the methodology reported[26] for fabrication of MA-devices under ambient conditions to formamidinium(FA)-devices by adjusting the DMSO proportion in precursor solution as a function of WVP. In particular, we fabricated $FA_{0.83}MA_{0.17}PbI_3$ at 1.56 kPa (302 K and 39 % RH). A planar configuration was chosen to simplify the devices manufacturing. Because DMSO coordinate worse for FA-perovskite than MA-perovskite,[53] we used a lower DMSO content in the solution precursor than that calibrated for $MAPbI_3$ (Fig. 2). By adjusting DMSO:$Pb^{2+}$ ratio to 0.75, FA-devices with an efficiency of 14.7 % ($J_{SC}$: 19.14 mA·$cm^{-2}$; $V_{OC}$: 1.03 V; FF: 74.7%) were obtained (Fig. 5). This efficiency is much higher than reported in the literature for FA-type of perovskite deposited at the same RH ( 9 % PCE at 40% RH).[19]

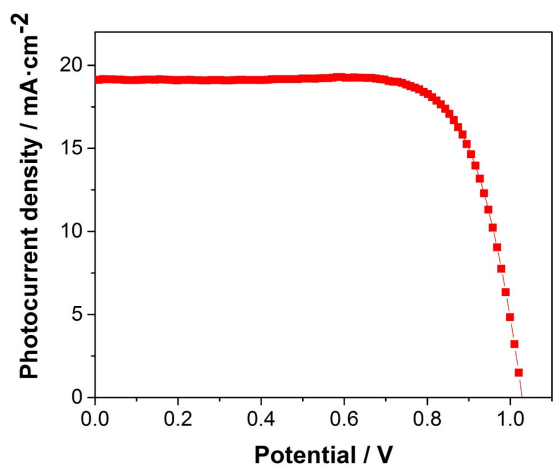

**Figure 5.** Current–voltage curve of the best FA$_{0.83}$MA$_{0.17}$PbI$_3$ devices prepared at 1.56 KPa (302 K and 39 % relative humidity) The curves have been measured in reverse scan under 1 sun—AM 1.5 illumination and using a mask of 0.16 cm². The table inset illustrates the photovoltaic parameters statistics for the 4 devices measured.

## Experimental details

### Fabrication of Perovskite solar cell

Perovskite solar cells were fabricated on FTO-coated glass (Pilkington–TEC15) patterned by laser etching. The substrates were cleaned using Hellmanex® solution and rinsed with deionized water and ethanol. Followed this they were sonicated in 2-propanol and dried by using compressed air. The TiO$_2$ blocking layer was deposited onto the substrates by spray pyrolysis at 450 °C, using a titanium diisopropoxide bis(acetylacetonate) solution (75% in 2-propanol, Sigma Aldrich) diluted in ethanol (1:14, v/v), with oxygen as carrier gas. The TiO$_2$ compact layer (c-TiO2) is kept at 450 ºC for 30 min for the formation of the anatase phase. Once the samples achieve room temperature, a TiO$_2$ mesoporous solution was deposited by spin coating at 2000 rpm during 10 s using a TiO$_2$ paste (Dyesol, 30NRD) diluted in ethanol (1:5, weight ratio) by strong stirring. After drying at 100 °C for 10 min, the TiO$_2$ mesoporous layer was sintered inside a furnace at 500 °C for 30 min and later cooled to room temperature. Subsequently, perovskite precursor solutions were prepared to be deposited at ambient conditions.

Pure methylammonium lead iodide (CH$_3$NH$_3$PbI$_3$, MAPbI$_3$) precursor solution with different DMSO:PbI$_2$ molar ratio (0; 0.5; 0.75; 1; 1.25; 1.5) were obtained from reacting DMF solutions (50 wt %) containing MAI and PbI$_2$ (1:1 mol %). While mixed cation perovskite ((CH(NH$_2$)$_2$)$_{0.83}$(CH$_3$NH$_3$)$_{0.17}$PbI$_3$, FA$_{0.83}$MA$_{0.17}$PbI$_3$) precursor solution was prepared from DMF solutions (50 wt %) containing FAI, MAI and PbI$_2$ (0.83:0.17:1 mol %) and molar ratio of DMSO and PbI$_2$ of 0.75 (mol %). The MAPbI$_3$ and FA$_{0.83}$MA$_{0.17}$PbI$_3$ precursor solution (50 µL) were deposited on mesoporous and compact TiO$_2$ layer, respectively, by spin coating in an one-step setup at 4000 rpm for 50 s. During this step, DMF is selectively washed with non-polar diethyl ether just before the white solid begins to crystallize in the substrate. Afterward the substrate was annealed at 100 °C for 3 min for MAPbI$_3$ and at 150 °C for 15 min for the FA$_{0.83}$MA$_{0.17}$PbI$_3$ (see Ref. 26 for details).

Spiro-OMeTAD was deposited as hole selective layer (HSL) by dissolving 72.3 mg in 1 mL of chlorobenzene as well as 17.5 µL of a lithium bis (trifluoromethylsulphonyl)imide (LiTFSI) stock solution (520 mg of LiTFSI in 1 mL of acetonitrile), and 28.8 µL of 4-tertbutylpyridine. The HTM solution was spin coated at 4000 rpm for 30 s. Finally, 60 nm of gold was deposited as a metallic contact by thermal evaporation under a vacuum level between $1 \cdot 10^{-6}$ and $1 \cdot 10^{-5}$ torr.

**Characterization of films and devices**

Current density–voltage (J–V) curves were documented with a Keithley 2400 source-measurement-unit under AM 1.5 G, 100 mW/cm$^2$ illumination from a 450 W AAA solar simulator (ORIEL, 94023 A). This was calibrated using a NREL certified calibrated mono-crystalline silicon solar cell. A metal mask was used to define an active area of 0.16 cm$^2$. The current-voltage curves were obtained using a reverse scan rate of 100 mV/s and sweep delay of 20s.

The illumination for the Impedance Spectroscopy (IS) measurements was provided by red ($\lambda$ = 635 nm) and blue ($\lambda$ = 465 nm) LEDs over a wide range of DC light intensities. This allows for probing the devices with two distinct optical penetrations.[36] A response analyzer module ((PGSTAT302N/FRA2, Autolab) was utilized to register the impedance spectra. In these experiments a 20 mV perturbation in the $10^6$ -0.1 Hz range was applied. To avoid voltage drop due to series resistance, the measurements were performed at the open circuit potential, the Fermi level (related to the open-circuit voltage) being fixed by the DC (bias) illumination intensity. To compensate for the different response under blue and red light due to the different optical absorption all parameters are monitored and plotted as a function of the open-circuit potential generated by each type of bias light. The NOVA 1.7 software was used to generate IS data and Z-view equivalent circuit modeling software (Scribner) for fitting the spectra.

UV-Visible absorption spectra were recorded by using a Cary 100 UV-Vis spectrophotometer (Agilent) in the range of 400-850 nm. Steady state photoluminescence measurements were performed using a Hitachi, F-7000 Fluorescence spectrophotometer. X-ray diffractograms were recorded on a Bruker-A25 (D8 Advance) diffractometer using a Cu-K$\alpha$1 (1.5406 Å) source. The diffractometer was set in *Grazing incidence geometry* at 2°. Scanning electron microscope (SEM) images of the samples were performed using a Zeiss GeminiSEM-300 microscope working at 2 kV

**Conclusions**

The PCE differences found for devices prepared at different room temperature for the same RH underlined the importance of using the water vapour pressure as the parameter that matters when controlling the humidity in the preparation of perovskite at ambient conditions. Taking into account WVP during perovskite deposition, a study of the effect of the absolute water content in air on the performance of devices was carried out. Simultaneously, the optimum DMSO content in precursor solution as a function of WVP was also analysed. Highest efficiencies were revealed for WVP values lower than 1.7 kPa and a DMSO:$Pb^{2+}$ ratio of 1.0. The best performance of devices was related with increased photocurrent and higher open-circuit potential. These results were attributed to faster electron injection into the $TiO_2$ layer. Higher electron recombination resistances were found in devices prepared at lower WVP, which seems to be related to the larger grain size. We demonstrated the broad applicability of a procedure based on the control of the absolute water content in air by fabricating double-cation perovskites ($FA_{0.83}MA_{0.17}PbI_3$) under humid conditions by adjusting the DMSO proportion in precursor solution as a function of WVP. Efficiencies higher than those reported for similar devices at ambient conditions were obtained. The results here reported show the potentiality of a method of preparation of stable and highly efficient perovskites devices at ambient conditions for industrial application as long as the *absolute* content level of water vapor do not surpass certain levels.

**Acknowledgments**

We thank Junta de Andalucía for financial support via grant FQM 1851 and FQM 2310, Ministerio de Economía y Competitividad of Spain under grants MAT2013-47192-C3-3-R and MAT2016-76892-C3-2-R and Red de Excelencia "Emerging photovoltaic

Technologies". We also thank "Servicio de Microscopía Electrónica de la Universidad Pablo de Olavide").

# Water vapour pressure as determining control parameter to fabricate high efficiency perovskite solar cells at ambient conditions – Electronic Suplementary Information

Lidia Contreras-Bernal[a], Juan Jesús Gallardo,[b] Javier Navas,[b] Jesús Idígoras[a],*, Juan A. Anta[a],*

[a] *Área de Química Física, Universidad Pablo de Olavide, E-41013, Sevilla, Spain*
[b] *Departamento de Química Física, Facultad de Ciencias, Universidad de Cádiz*, E-11510 Puerto Real (Cádiz), Spain

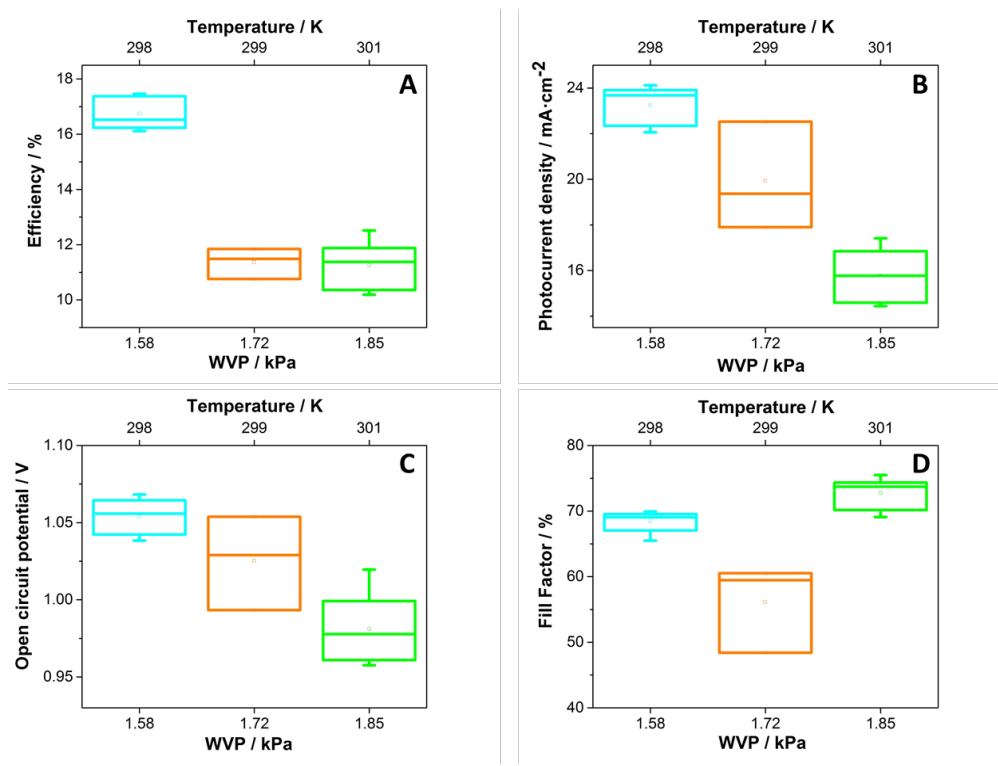

**Figure S1**. Photovoltaic parameters statistics using analysis of variance (ANOVA) of MAPbI$_3$ devices prepared under different temperature conditions and a relative humidity of 50%. Data were obtained in reverse scan under AM1.5 - 1 sun illumination for perovskite films deposited from a solution precursor with a ratio DMSO:Pb$^{2+}$ of 0.75. Note that at least 6 devices of each configuration were measured.

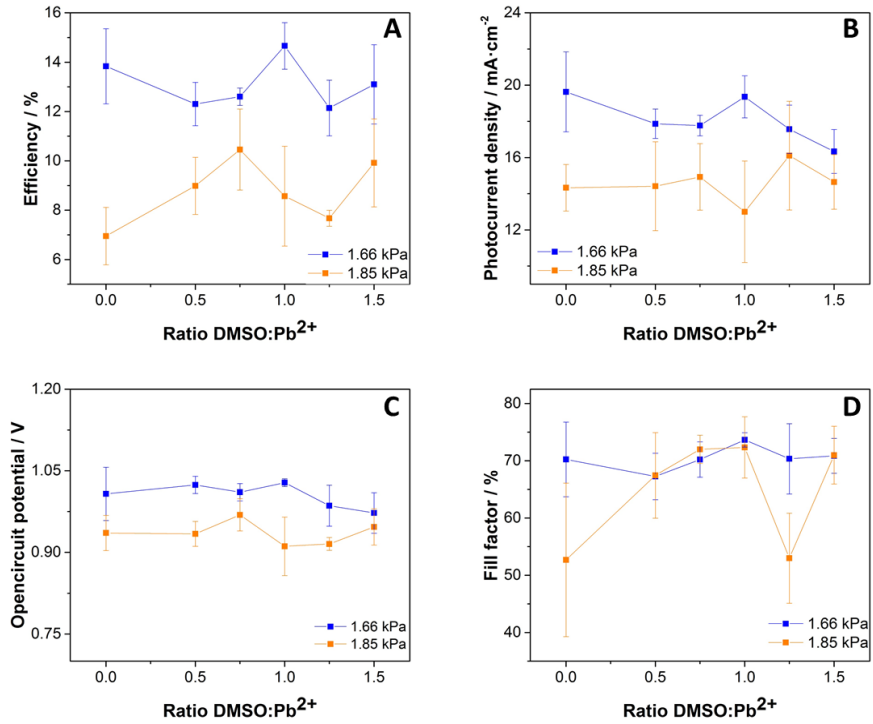

**Figure S2**. Photovoltaic parameters of MAPbI$_3$ devices prepared from precursor solution with different DMSO:Pb$^{2+}$ ratio for two water vapour pressure (WVP). Data were obtained in reverse scan under AM1.5 - 1 sun illumination for perovskite solar devices fabricated at 301 K and 44 % and 49 % R.H (1.66 kPa and 1.85 kPa, respectively).

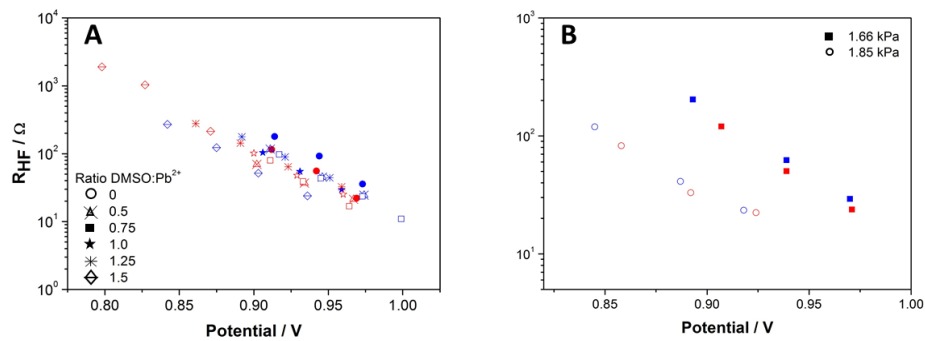

**Figure S3**. Electron recombination resistance data as extracted from impedance spectroscopy measurements for MAPbI$_3$ devices prepared (A) from precursor solution with different DMSO:Pb$^{2+}$ ratios and deposited under 1.66 kPa and (B) for two different water vapour pressure (WVP) for ratio 0.75. Results using excitation wavelengths 635 nm (blue symbols) and 465 nm (red symbols) are shown.

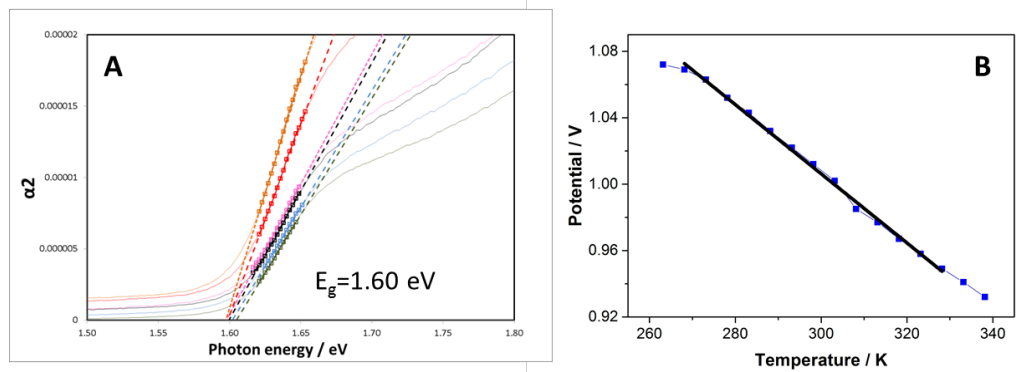

**Figure S4.** (A) Illustration the estimation of the optical bandgap from the measured spectra for MAPbI$_3$ devices prepared from precursor solution with different DMSO:Pb$^{2+}$ ratio (B) Open-circuit potential as a function of temperature for MAPbI$_3$ devices from DMSO:Pb$^{2+}$ ratio 1.0 for white light and a light intensity of 14.15 W/m$^2$. Date obtained for devices prepared under 1.66 kPa.

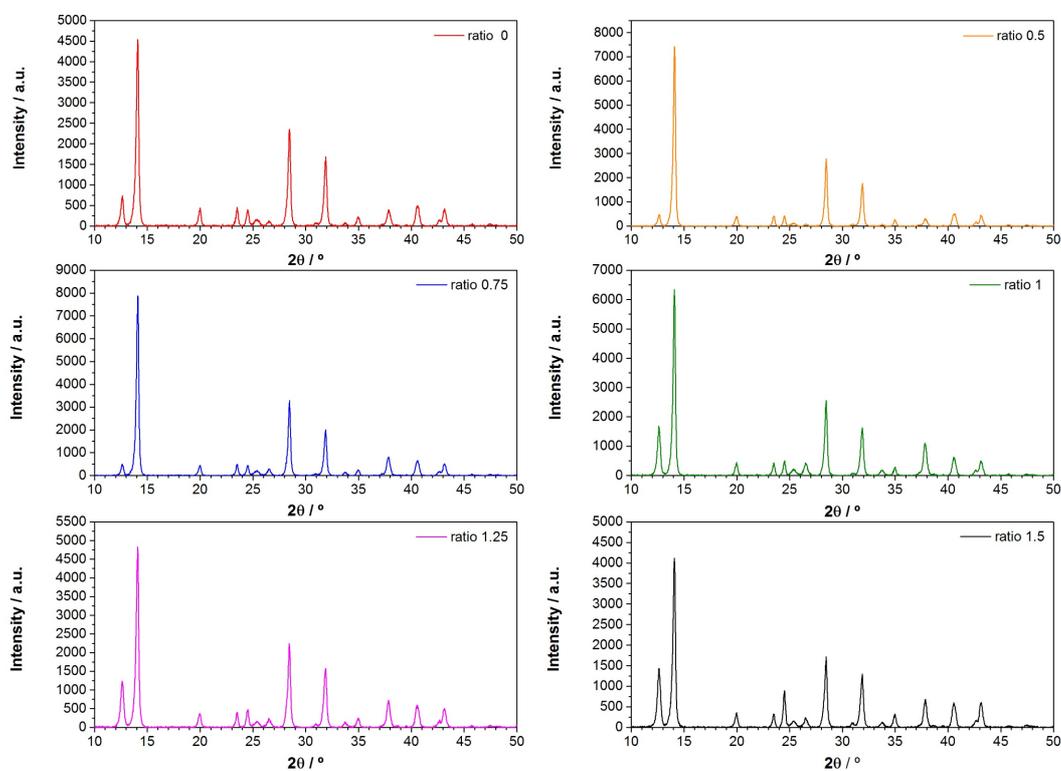

**Figure S5.** XRD pattern of composite MAPbI$_3$/mesoporous-TiO$_2$ films prepared under 1.01 kPa (299 K and 30 % relative humidity)

**Table S1.** PbI$_2$/ MAPbI$_3$ ratio for films deposited under 1.01 KPa (299 K and 30 % relative humidity)

| Ratio DMSO:Pb$^{2+}$ | PbI$_2$ | | MAPbI$_3$ | | I$_1$/I$_2$ |
|---|---|---|---|---|---|
| | 2θ | I$_1$ | 2θ | I$_2$ | |
| 0 | 12.65 | 740.9 | 14.08 | 4539.6 | 0.163 |
| 0.5 | 12.65 | 472.4 | 14.1 | 7424.5 | 0.064 |
| 0.75 | 12.63 | 498.7 | 14.1 | 7881.7 | 0.064 |
| 1 | 12.61 | 1690.6 | 14.08 | 6343.2 | 0.266 |
| 1.25 | 12.61 | 1241.8 | 14.08 | 4829.2 | 0.257 |
| 1.5 | 12.63 | 1438.4 | 14.1 | 4114.9 | 0.35 |

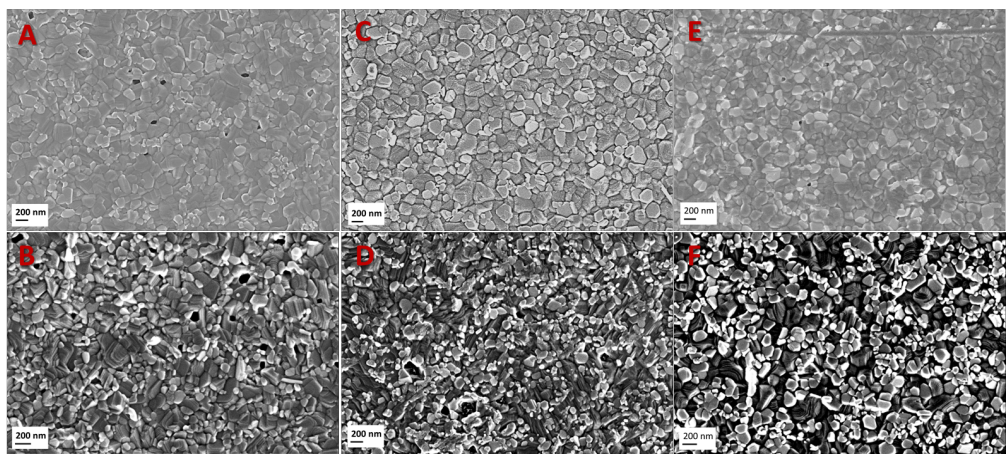

**Figure S6.** Top-view scanning electron microscopy (SEM) images for MAPbI$_3$ films from precursor solution with DMSO:Pb$^{2+}$ ratio 1.5 (A and B), 1.0 (C and D) and 0.5 (E and F) and deposited at 1.66 kPa (A, C and E) and 1.85 kPa (B, D and F).